\documentstyle[aps,prl,preprint]{revtex}

\tighten
\include{psfig}
\begin{document}
\draft \title{\bf Inherent Structure Entropy of Supercooled Liquids}

\author{F. Sciortino$^{(1)}$,  W. Kob$^{(2)}$ and P. Tartaglia$^{(1)}$ }

\address{$^{(1)}$ Dipartimento di Fisica and Istituto Nazionale
 per la Fisica della Materia, Universit\'a di Roma {\it La Sapienza},\\
 P.le Aldo Moro 2, I-00185, Roma, Italy \\
$^{(2)}$ Institut f\"ur Physik, Johannes Gutenberg--Universit\"at,
Staudinger Weg 7, D-55099 Mainz, Germany }

\date{\today} \maketitle

\begin{abstract}
We present a quantitative description of the thermodynamics 
in a supercooled binary Lennard Jones liquid via the
evaluation of the degeneracy of the inherent structures, i.e. of the
number of potential energy basins in configuration space. 
We find that  for
supercooled states, the contribution 
of the inherent structures to the free energy of the liquid 
almost completely  decouples from 
the vibrational contribution. An important byproduct of the
presented analysis is the determination of the Kauzmann temperature for
the studied system. 
The resulting quantitative picture 
of the thermodynamics of the inherent structures offers 
new suggestions for
the description of equilibrium and out-of-equilibrium 
slow-dynamics in liquids below
the Mode-Coupling temperature.

\end{abstract}

In recent years, a significant effort has been devoted to
understanding the fundamental nature  of glass-forming materials, 
a long-standing open problem in condensed matter physics\cite{review-glass}. 
Theoretical\cite{goetze,parisi,speedy}, 
experimental\cite{review-exp,angell-science} 
and numerical efforts\cite{kob-review}
have broadened our knowledge of the
physical mechanisms  responsible for the dramatic slowing down of the
dynamics in supercooled liquids 
(more than 15 order of magnitudes) as the temperature $T$ changes 
over a small range, as the glass transition temperature is approached.

Some recent theoretical approaches build upon ideas
which were presented several decades 
ago\cite{kauzman,adam-gibbs,goldstein,dimarzio,sw}. In these works,
the slowing down of the dynamics was connected to the presence of
basins in configuration space.
The short time dynamics
(on a $ps$ time scale) was related to the process of 
exploring a finite region of phase space around a local potential energy 
minimum, while
the  long time dynamics was connected to the transition among 
different local potential energy minima\cite{goldstein,sw}.
In this picture, upon  cooling    
the 
intra-basin motion becomes more and more separated 
in time from the slow (and strongly $T$-dependent) 
inter-basins motion.
The decrease
of the entropy of supercooled liquids on cooling\cite{kauzman} 
was associated with
the progressive ordering of the system in configuration space, i.e.
in the progressive population of  
basins with deeper energy  but of lower degeneracy\cite{adam-gibbs}.

Following these ideas, Stillinger and Weber\cite{sw}
introduced the concept of inherent structure (IS), defined as
{\it local} minimum configuration of the
$3N$-dimensional potential energy surface. A basin 
in configuration space was defined as the set of points that 
 --- via a steepest descent path along the potential energy 
hypersurface ---  map to the same 
IS. 
This precise operational
definition of a basin allows the 
configuration space to be partitioned into  an
ensemble of distinct basins. Thus, the 
canonical partition function $Z_N$ for a system of $N$ atoms
at inverse temperature 
$\beta=1/k_BT$  can be written as

\begin{equation}
Z_{N}=\lambda^{-3N} \sum_{\alpha} \exp(-{\beta} \Phi_{\alpha})
\int_{R_{\alpha}} \exp [-\beta \Delta_{\alpha} ({\bf r}^N)]d {\bf r}^N
\label{eq:zeta1}
\end{equation}

\noindent         
where $R_{\alpha}$ is the set of points composing the basin $\alpha$, 
$\Phi_{\alpha}$ is the potential energy of minimum $\alpha$
and the non-negative quantity $ \Delta_{\alpha} ({\bf r}^N)$  
measures the potential energy at a point ${\bf r}^N$ 
belonging to the basin $\alpha$ relative to the minimum. 
The integration over the
momenta introduces the thermal wavelength  
$\lambda= \sqrt{ \beta h^2 /2 \pi m}$, where $m$ is the mass. 
Eq. (\ref{eq:zeta1}) shows that  both the IS energy 
and  the thermal  excitation
within the basin region $R_{\alpha}$ contribute to $Z_N$. 
Stillinger and Weber also noted that, 
if the value of the  potential energy  minimum uniquely characterizes the
properties of the basin, then a very powerful simplification of 
Eq. (\ref{eq:zeta1}) can be performed.  By introducing a density
of states $\Omega(e_{IS})$  with IS-energy $e_{IS}$, $Z_N$ can be written as

\begin{equation}
Z_{N}  \approx 
\int de_{IS} \Omega(e_{IS}) 
\exp [-{\beta} e_{IS}- \beta f(\beta, e_{IS})]
\label{eq:zeta2}
\end{equation}

\noindent
where 
\begin{equation}
- \beta f(\beta, e_{IS}) = ln (  
\int_{R(e_{IS})}  \exp [-\beta \Delta_{e_{IS}} \Phi({\bf r}^N)]~ {{d{\bf r}^N} 
\over {\lambda^{3N}}}  ) 
\label{eq:freeenergy}
\end{equation}

\noindent
can be interpreted as free energy of the system 
when confined in
one of the characteristic basins with IS energy $e_{IS}$. 
Then, the probability that a
configuration of the liquid extracted from an equilibrium
ensemble of configurations at temperature $T$ is associated to 
an IS with energy between $e_{IS}$ and $e_{IS} + \delta e_{IS}$  is

\begin{equation}
P(e_{IS},T)= {{ \Omega(e_{IS}) \delta e_{IS} 
exp [-{\beta} e_{IS}- \beta f(\beta, e_{IS})] }\over {Z_N(\beta)}}
=
{
{ exp [-{\beta} (  e_{IS}- T S_{conf}(e_{IS}) +  f(\beta, e_{IS}))] } 
\over {Z_N(\beta)} }
\label{eq:pofe}
\end{equation}

\noindent
where we have defined $ S_{conf}(e_{IS}) \equiv k_B 
ln [\Omega(e_{IS})\delta e_{IS}    ]$, since $\Omega(e_{IS}) \delta e_{IS} $ is the number of states between $e_{IS}$ and $e_{IS} + \delta e_{IS}$.

The formalism proposed by Stillinger and Weber, 
although often used in the past to clarify 
structural issues in liquids\cite{useofsw,water}, 
for long time was not quantitatively 
applied to computer studies of the glass-transition problem, 
due to the significant computational effort required in equilibrating
atomic configurations  at low $T$. 
Only recently, Sastry {\it et al} \cite{nature} 
addressed the problem of evaluating the $T$ dependence of
the average IS energy ($\bar e_{IS}$) in supercooled states 
in a binary mixture of LJ particles,  
observing a significant decrease of  $\bar e_{IS}$ 
on supercooling.
This result, also observed for 
other models for liquids, e.g. in models for water\cite{water} and
orthoterphenyl\cite{ortho},  furnishes strong evidence of the relevant role
played by the low-energy basins on cooling.  In a recent work\cite{aging}, 
we proposed 
to invert the relation between the $\bar e_{IS}$ energy and $T$  
to define an effective temperature 
at which the configurational part of the system is in equilibrium.
This hypothesis, which has been proven useful in interpreting the
aging process in a model liquid in terms of progressive thermalization
of the IS\cite{aging}, support the validity of 
Eq. (\ref{eq:zeta2}) and
together with the
work of Sastry et al., calls for an
effort in the direction of checking the formal expression
for the supercooled liquid free energy, i.e. the $T$-range 
of validity, and an effort in the direction of evaluating
the $e_{IS}$ dependence of the configurational entropy. 
This article is a first effort in this direction.

The model system we study is the well-known 80-20 Lennard Jones
$A-B$ binary mixture\cite{kob}, 
composed by a 1000 atoms in a volume $V_o=(9.4)^3$,
corresponding to a reduced density of 1.2039. Units of length and
energy are defined by the   $\sigma$ and $\epsilon$  parameters of the 
$A-A$  Lennard Jones interaction potential. The mass of atom $A$ is chosen 
to be 1. In these units, $k_B=1$.
Simulations, covering the range $0.446 < T < 5 $, 
have been performed in 
the canonical ensemble by coupling the system to 
a Nose'-Hoover thermostat\cite{nh}. 
This system is well characterized and its slow dynamics has been
studied extensively\cite{kob}. The
critical temperature of Mode Coupling Theory $T_{MCT}$ 
for this system is $0.435$\cite{kob}.

Between 500 and 1000 equilibrium configurations for each $T$
(covering more than $80$ millions integration time steps for each $T$ )
have been  quenched to their local minima by using 
a standard conjugate-gradient minimization algorithm. 
By performing this large number of quenches we are able to 
determine 
not only $\bar e_{IS}$
and its $T$ dependence but also the
probability distribution $P(e_{IS},T)$, shown in Fig.\ref{fig:fig1}-(A).  

In the $T$-region where Eq. (\ref{eq:pofe}) is supposed to hold,
curves of $ln[P(e_{IS},T)]+ \beta e_{IS}$  
are equal to  $S_{conf}(e_{IS})/k_B - 
\beta f(\beta,e_{IS})$,  except for 
the $T$-dependent constant $ln[Z(\beta)]$. If  
$f(\beta,e_{IS})$  has only a weak dependence on  $e_{IS}$, then 
it is possible to 
superimpose  $P(e_{IS},T)$ curves at different temperatures
which overlap in $e_{IS}$. The resulting $e_{IS}$-dependent curve
is, except for an unknown constant, $S_{conf}(e_{IS})/k_B$ 
in the $e_{IS}$ range sampled within the
studied $T$ interval.
This procedure is displayed in Fig.\ref{fig:fig1}-(B).
We note that while
below $T=0.8$ curves for different $T$ lie on the same master curve, 
above $T=0.8$, curves for different $T$ have different $e_{IS}$
dependence, thus showing the progressive $e_{IS}$-dependence of
$f(\beta,e_{IS})$. The overlap between different $P(e_{IS},T)$ curves
below $T=0.8$ indicates that the obtained master curve is indeed, except for an unknown constant,
the $e_{IS}$ configurational entropy, i.e. the logarithm of  
the number of basins with the same $e_{IS}$ value.

It is particularly relevant that, for $T < 0.8$, 
$f(\beta,e_{IS}) \approx f(\beta)$ and thus
$Z_N$  (Eq. \ref{eq:zeta2}) is well approximated  by
the product of a vibrational contribution  [$ e^{- \beta f}$] 
and of a configurational contribution depending only on
the IS-energies and their degeneracy [$\int de_{IS} \Omega(e_{IS})
e^{- \beta e_{IS}} $]. Thus the liquid can be considered as composed by two
independent subsystems, respectively described by the
IS and by the vibrational part. The IS subsystem can be considered 
as a continuum of levels characterized by
an energy value $e_{IS}$ and an associated degeneracy $\Omega(e_{IS}) $.
When the IS-subsystem is in thermal equilibrium
(with the vibrational subsystem), then the $T$ dependence of the average 
configurational entropy $\bar S_{conf}$
can be evaluated using the standard thermodynamic relation  

\begin{equation}
{{ d \bar{S}_{conf}(T)} \over {d \bar{e}_{IS}}}=   {{1}\over{T}}
\label{eq:sdie}
\end{equation}

\noindent
i.e. by integrating the $T$ dependence of $d\bar e_{IS} /T$.

While the evaluation of the $T$-dependence of $S_{conf}$ already furnishes
relevant information, the evaluation of the unknown integration 
constant  would allow for a determination of the
number of IS  with the same $e_{IS}$ and, via a suitable
low $T$ extrapolation, to the determination of the so-called 
Kauzmann temperature $T_K$, i.e. the $T$ at which the
configurational entropy appears to approach zero.
To do so we exploit the fact that Eq. \ref{eq:zeta2} predicts that
the liquid free energy $F_{liquid}(T)$ can be written as\cite{saddle}

\begin{equation}
F_{liquid}(T) =  -k_B T ln[Z_N(T)] =  \bar e_{IS}(T)- T S_{conf}(\bar e_{IS}(T)) + 
f(\beta, \bar e_{IS}(T)) 
\label{eq:fel}
\end{equation}

\noindent
where $\bar e_{IS}(T)$ is the $e_{IS}$ value which maximises the 
integrand in Eq. (\ref{eq:zeta2}).
We assume that at the lowest studied $T$, 
the unknown  $f(\beta, \bar e_{IS})$ can be approximated 
by the harmonic free energy of a
disordered system characterized by the eigenfrequencies spectrum
calculated from the distribution of IS at
the corresponding $T$\cite{harmonic}. 
In this approximation,  the entropy of the liquid,
which can be calculate via thermodynamic integration, once the
entropy of the corresponding harmonic disordered solid is  subtracted, 
provides an estimate of
the configurational entropy in absolute units.

To calculate the liquid entropy 
we perform a thermodynamic integration first along the $T=5.0$ isotherm, 
from infinite volume  down to $V_o$, followed
by  a $T$ integration of the specific heat 
at fixed volume, down to the lowest studied temperature.
Since the $T$ dependence of the potential energy $E$ along the studied 
isochore is extremely well
described by the law $E(T) \sim  T^{3/5}$,  
in agreement with recent theoretical predictions 
for dense fluids\cite{talla35},
the  potential energy contribution to the liquid entropy 
follows the law $ T^{-2/5}$. By summing the  
kinetic energy contribution we obtain an analytic expression
which can reliably be extrapolated to temperatures lower than the
studied ones.

The evaluated $T$ dependence of liquid and disordered-solid entropies
is reported in Fig.\ref{fig:fig2}. We also show in Fig. \ref{fig:fig2} the
contribution to the
disordered-solid entropy arising from the $T$-dependence of the 
eigenfrequencies spectrum, to
confirm that the $T$-dependence of the harmonic solid frequencies 
contribute only weakly to the entropy.
The $T$ dependence of $ \Delta S \equiv S_{liquid}$-$S_{disordered-solid}$ 
is shown in Fig.\ref{fig:fig3}-(A).
We note that this difference
vanishes at $T=0.297 \pm 0.02 $, which defines $T_K$
for the studied binary mixture. An independent recent estimate of
$T_K$ for this system, based on an integral equation approach and on a 
similar analysis of simulation data,  
is $T_K=0.29$ \cite{barbara}.
The resulting ratio between  $T_K$ and  $T_{MCT}$   
support the view that the studied system has
intermediate fragility character, as recently predicted by Angell and
coworkers on the basis of a comparison between 
experimental results and numerical data for the
same system\cite{austen-pisa}.

At the lowest studied temperatures, where the harmonic approximation
is valid\cite{harmonic}, $ \Delta S $ coincides
with $S_{conf}(T)$. We use this identity to calculate the
unknown integration constant for the inherent structure entropy.
Moreover, by integrating $T d (\Delta S (T)) $,
both the configurational energy dependence of the
configurational entropy (Fig. \ref{fig:fig1}-(c)) and
the $T$ dependence of the configurational energy  (\ref{fig:fig3}-(b))
can be evaluated, allowing to
bridge the gap between $T_K$ and the lowest $T$ at which we were able to
equilibrate the system. 
The present analysis predicts  $e_{IS}(T_K)= -7.82 \pm 0.01$. 
Thus, both the configuration entropy and energy 
around  $T_{MCT}$ is halfway between $T_K$ and the
high $T$ value, suggesting that the ordering process in configuration space
at the lowest  temperature which we have been able to equilibrate is far 
from being complete.  

The data reported in this article offer a quantitative
thermodynamic analysis of the supercooling state. This picture
confirms the fruitful ideas put forward long time 
ago\cite{kauzman,adam-gibbs,goldstein,dimarzio,sw} and shows
that a thermodynamic
approach for the inherent structures subsystem becomes 
possible in supercooled states, 
since  the thermodynamics of the  inherent structures  
almost completely decouples
from the ``vibrational'' thermodynamics.
In particular, the
quantitative evaluation of the 
degeneracy of the inherent structures  for a well characterized system
constitutes a basis for a comprehensive
description of the slow dynamics below $T_{MCT}$, the $T$-range in which 
an accurate theoretical prediction is still missing.
Previous estimates of the
degeneracy of $e_{IS}$, the key ingredient for a descriptions of the
dynamics in terms of IS, were available only for
systems composed by less than 50 atoms\cite{heuer,viliani}.
The proposed view of a supercooled liquids as composed by two weakly coupled
subsystems --- the IS subsystem and the ``vibrational'' subsystem ---
will also offer stimulating ideas for a microscopic understanding both of the
out-of-equilibrium thermodynamics theories recently proposed\cite{teo} 
and of the aging process\cite{aging,cugliandolo}.

Acknowledgment:
We thank  B. Coluzzi, G. Parisi and P. Verrocchio for sharing with us
their recent independent results of the BM LJ system (\cite{barbara}).
We thank G. Parisi for bringing to our
attention Ref.\protect\cite{talla35}. We thank S. Ciuchi, A. Crisanti,
K. Kawasaki, E. La Nave, P. Poole, A. Scala
for comments. F.S. and P.T. acknowledge support from MURST PRIN 98 and W.K.
from the DFG through SFB~262.

\begin{figure}


\caption{ (A) 
Distributions $ P(e_{IS},T)$  of the $IS$ energy (per atom)
for different equilibrium temperatures $T$.  From left to right:
$T=0.446, 0.466, 0.5, 0.55, 0.6, 0.8, 1.0, 2.0, 4.0$. 
(B)  $ln[P(e_{IS},T)]+ \beta e_{IS} +C(T) $, 
for different equilibrium temperatures $T$.  
Symbols refer to different $T$-values.
The curves which do not lay on the continuous line correspond to
$T=4.0,2.0,1.5,1.0,0.8$, from bottom to top. 
The constant $C(T)$ 
has been chosen to maximize the overlap between
curves with different $T$ and the overlap with 
$S_{conf}(e_{IS})$ (in absolute units) 
calculated  from $\Delta S(T)$, as discussed in the text. 
$S_{conf}(e_{IS})$ is reported as continuous line both in (B) 
and in the enlarged $e_{IS}$ range
in (C). Note that when $e_{IS}=-7.82$, $S_{conf}=0$.}
\label{fig:fig1}
\end{figure}

\begin{figure}

\caption{Disordered-solid (filled squares) and liquid (filled circles) 
entropies as a function of $T$.
Filled rombs are  $S_{disordered-solid}- 3 N ln (T/T_0)$, with $T_0=1$, 
to demonstrate the  
weakness of the $T$ contribution arising from the $T$-dependence of frequency
distribution. Below $T=1$, this weak $T$-dependent contribution 
is well fitted (long-dashed line) by the quadratic
polynomial $ 3201.62309-402.760885 T + 199.228407 T^2$,  
providing a reliable extrapolation of $S_{disordered-solid}$ 
to $T$ lower than the studied ones (dotted line). The liquid
entropy is extrapolated according to the theoretical predictions
discussed in Ref.\protect\cite{talla35}, i.e. 
$S_{liquid} + 1.5 k_B ln(T/T_0) \sim T^{-2/5}$
}
\label{fig:fig2}
\end{figure}

\begin{figure}


\caption{
Configurational entropy (A)  and energy (B) 
as a function of $T$ calculated as difference
of the liquid and disordered-solid entropies and energies, respectively. 
The arrow indicates 
$T_{MCT}$ for this system. The full circles in (B) are 
$\bar e_{IS}(T)$. Note that in the region where the
harmonic approximation for the disordered solid is expected to be valid
\protect\cite{harmonic}, $\bar e_{IS}(T)$ coincides with the configurational
energy.}
\label{fig:fig3}
\end{figure}

\setcounter{figure}{0}

\begin{figure}
\centerline{\psfig{figure=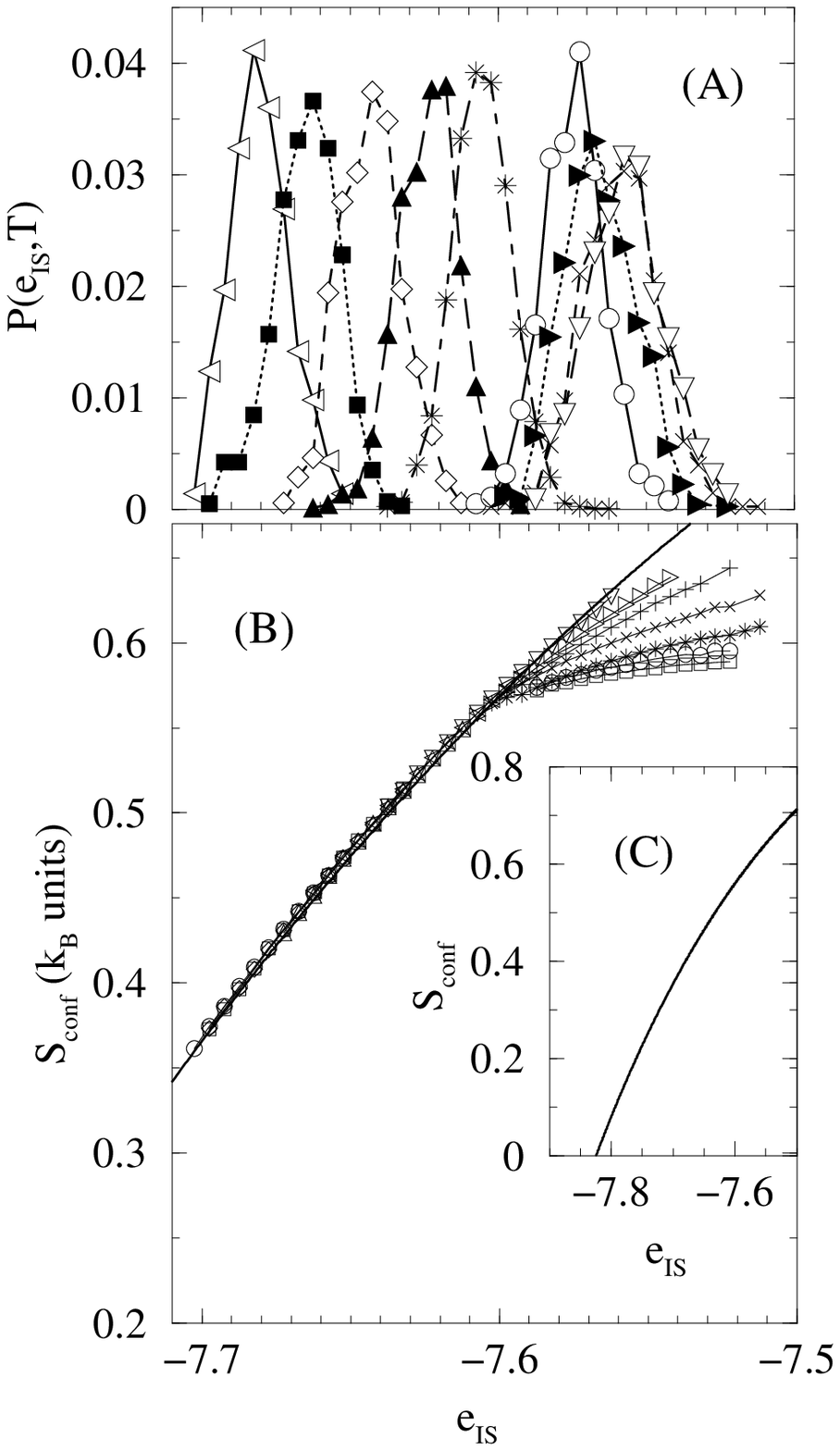,height=17.5cm,width=11.cm,clip=,angle=0.}}
\caption{ F. Sciortino et al}
\end{figure}

\begin{figure}
\centerline{\psfig{figure=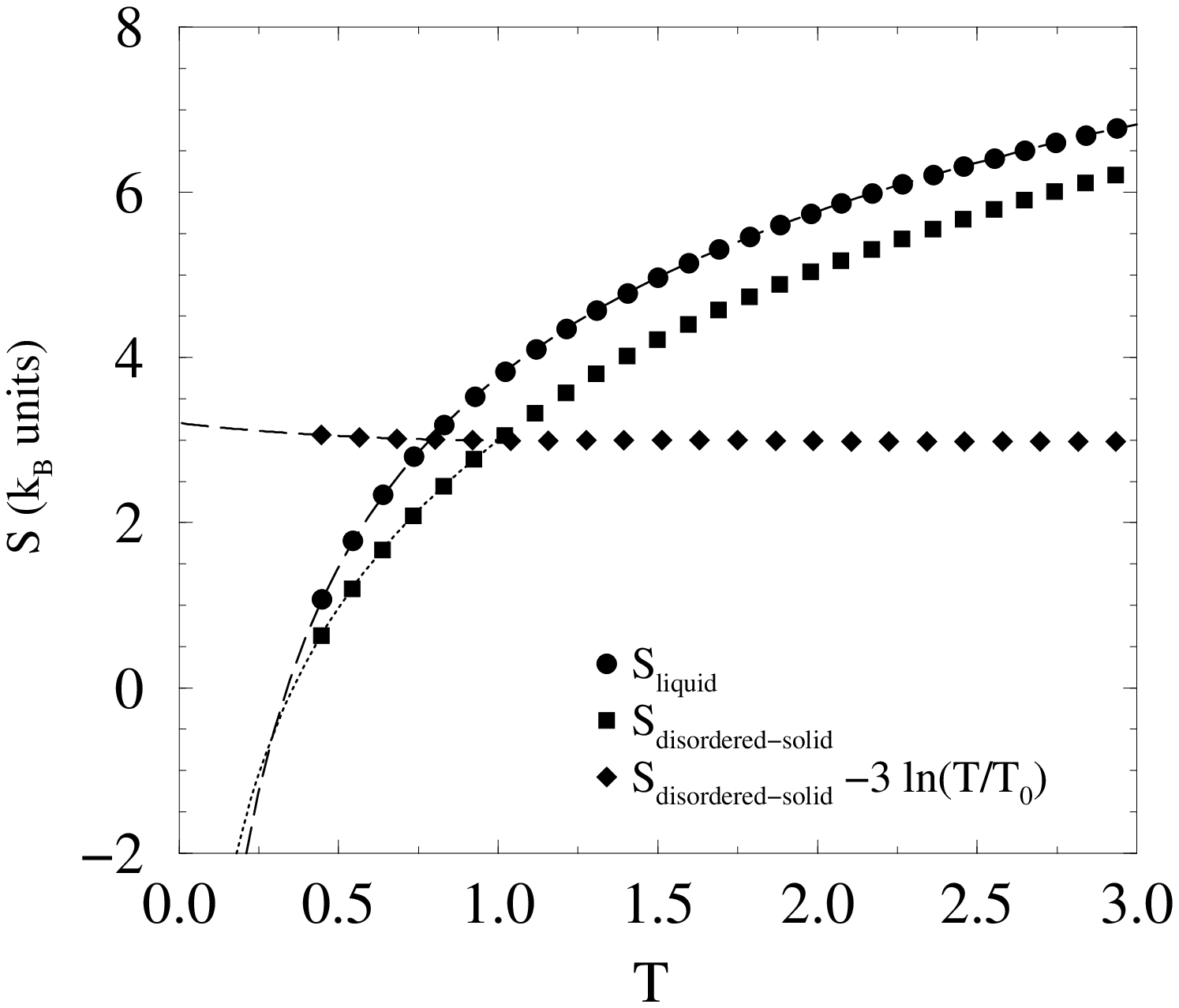,height=14cm,width=16cm,clip=,angle=0.}}
\caption{F. Sciortino et al}
\end{figure}

\begin{figure}
\centerline{\psfig{figure=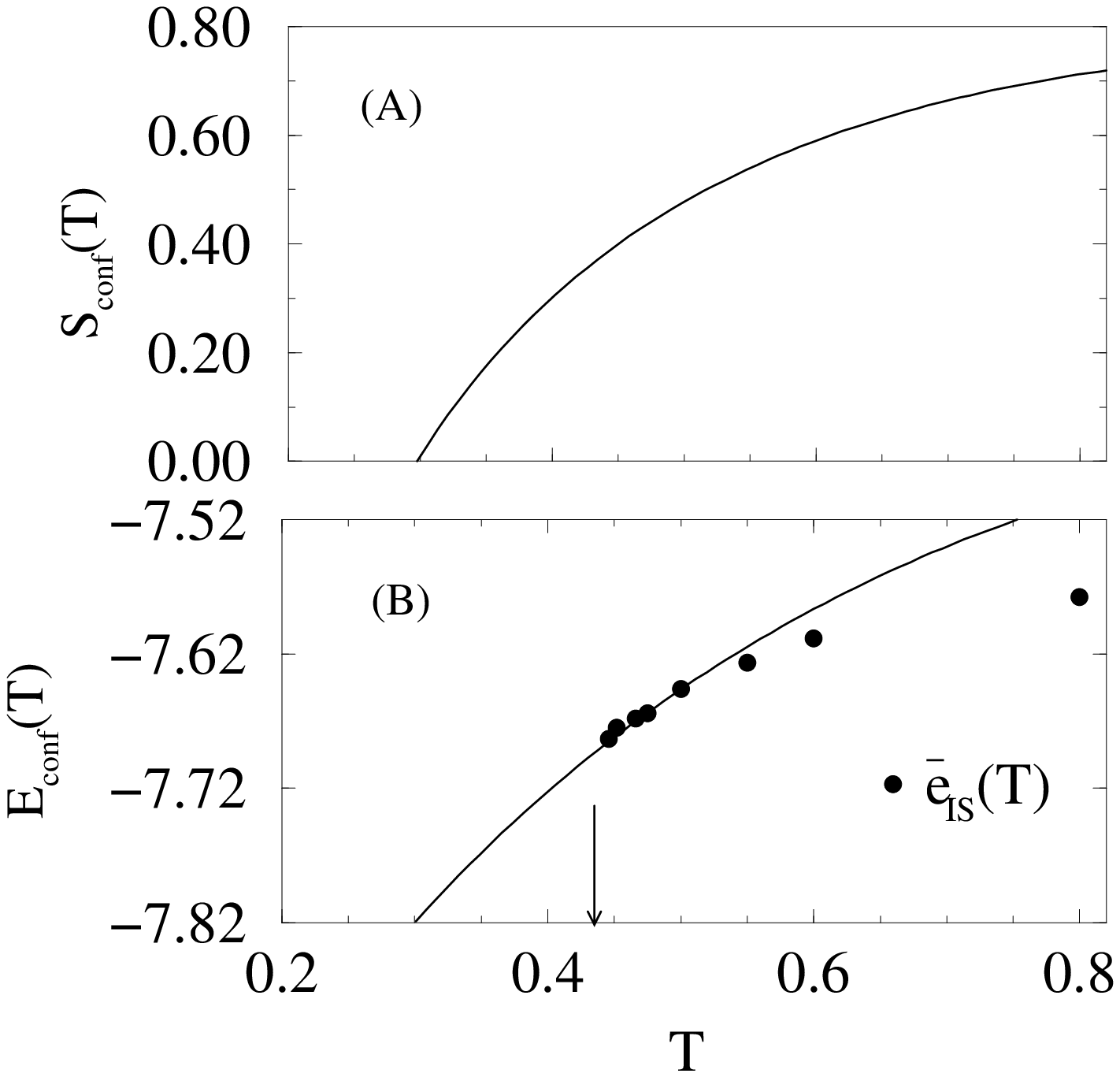,height=16cm,width=15cm,clip=,angle=0.}}
\caption{F. Sciortino et al}
\end{figure}


\begin{references}
\bibitem{saddle} In going from Eq.(\protect\ref{eq:zeta2}) to 
 Eq.(\protect\ref{eq:fel}) we have performed a maximum integrand 
evaluation of the
partition function, a procedure which becomes 
exact in the thermodynamic limit.
\bibitem{harmonic} Below $T=0.5$, $e_{IS}(T)+{ {3} \over {2} } k_BT$
is equal to the average potential energy, supporting the hypothesis that
the system mainly probes an harmonic potential landscape.
\end{references}
\end{document}